\documentclass[useAMS,usenatbib]{mn2e}

\title[Propagation of gamma-rays in LS 5039 and LSI +61$^{\rm o}$ 303]
{Propagation of  very high energy gamma-rays inside massive binaries 
LS 5039 and LSI +61$^{\rm o}$ 303}
\author[W. Bednarek]{W. Bednarek\thanks{E-mail:
bednar@fizwe4.fic.uni.lodz.pl}\\
Department of Experimental Physics, University of \L \'od\'z,
ul. Pomorska 149/153, 90-236 \L \'od\'z, Poland}
\begin{document}

\date{Accepted . Received ; in original form }

\pagerange{\pageref{firstpage}--\pageref{lastpage}} \pubyear{2005}

\maketitle

\label{firstpage}

\begin{abstract}

Two massive binary systems of the microquasar type, LS 5039 and LSI +61$^{\rm o}$ 303,
have been suggested as possible counterparts of EGRET sources. LS 5039
has been also recently detected in TeV gamma-rays. Since the massive stars in these
binary systems are very luminous, it is expected that high energy gamma-rays, if 
injected relatively close to the massive stars, should be strongly absorbed, 
initiating inverse Compton $e^\pm$ pair cascades in the anisotropic radiation from 
stellar surfaces. We investigate influence of the propagation effects on
the spectral and angular features of the 
$\gamma$-ray spectra emerging from these two binary systems by applying the Monte Carlo
method. Two different hypothesis are considered: 
isotropic injection of primary gamma-rays with the power law spectrum due to e.g. 
interaction of hadrons with the matter of the wind, and the isotropic injection of 
electrons, e.g. accelerated in the jet, which comptonize the radiation from the massive star.
It is concluded that propagation effects of $\gamma$-rays
can be responsible for the spectral features observed
from LS 5039 (e.g. the shape of the spectrum in the GeV and TeV energy ranges and 
their relative luminosities). The cascade processes occurring inside these
binary systems significantly reduce the gamma-ray opacity obtained in other works
by simple calculations of the escape of gamma-rays from the radiation fields 
of the massive stars.  Both systems provide very similar
conditions for the TeV $\gamma$-ray production at the periastron passage.
Any TeV $\gamma$-ray flux at the apastron passage in LSI +61$^{\rm o}$ 303 
will be relatively stronger with respect to its GeV flux than in  
LS 5039. If $\gamma$-rays are produced inside these binaries not far from 
the massive stars, i.e. within a few 
stellar radii, then clear anticorrelation between the GeV and TeV 
emission should be observed, provided that primary $\gamma$-rays
at GeV and TeV energies  
are produced in the same process by the same population of relativistic particles.
These $\gamma$-ray propagation features 
can be tested in the near future by the multi-wavelength campaigns 
engaging the AGILE and GLAST telescopes ($>$30 MeV) and the Cherenkov 
telescopes ($>$100 GeV, e.g. MAGIC, HESS, VERITAS and CANGAROO).
\end{abstract}

\section{Introduction}

Massive binary systems provide very promising conditions for acceleration of 
particles to relativistic energies and also well defined conditions 
for their possible interaction. 
Therefore, they have been considered for a long time as possible 
sources of high energy $\gamma$-rays and neutrinos in which  
acceleration processes can be tested. In fact, some GeV $\gamma$-ray 
sources observed by EGRET detector ($>100$ MeV) have been proposed
to be related to well known massive binaries in which non-thermal processes
were evident at lower energies,
e.g. LSI +61$^{\rm o}$ 303 (2EG J0241+6119; Thompson et al.~1995), Cen X-3 
(Vestrand, Sreekumar \& Mori~1997), Cyg X-3 (2EG J2033-4112; Mori et al.~1997), 
and LS 5039 (3EG J1824-1514; Paredes et al.~2000).
Early searches of the TeV $\gamma$-ray signals from massive binaries have not been  
very convincing, with the exception of Cen X-3 binary system 
(containing slowly rotating neutron star) 
which has been reported as TeV $\gamma$-ray source by the Durham group
(Chadwick et al.~1998, 1999, Atoyan et al.~2002).
The turning-point came recently with the observations of TeV $\gamma$-ray signals 
from two massive binaries: PSR B1259-63/SS 2883, containing a radio pulsar with 
the period of 47.8 ms (Aharonian et al.~2005a), and LS 5039, so called microquasar 
possibly containing a solar mass black hole (Aharonian et al.~2005b).   

The $\gamma$-ray emission from massive binaries is usually interpreted in terms of
the inverse Compton scattering (ICS) model in which thermal radiation coming from the
stellar surface is scattered by electrons accelerated in the pulsar wind shock
(e.g. Maraschi \& Treves~1981) or by electrons moving highly anisotropically
in the form of beams or jets (e.g. Bednarek et al.~1990). 
The binary systems which are considered as an example in this paper, 
LS 5039 and LSI +61$^{\rm o}$ 303, have been recently discussed in terms of the
microquasar IC model by Bosch-Ramon \& Paredes~(2004a,b) and Paredes, Bosch-Ramon \& 
Romero~(2005). Primary $\gamma$-rays might be also produced in the microquasar 
scenario of the binary system as a result of the interaction of hadrons, accelerated
in the jet, with the matter of the stellar wind (Romero, Christiansen \& Orellana~2005).
For review of the $\gamma$-ray production in microquasars see e.g. Romero (2004) or
Paredes~(2005).
  
The problem of importance of the propagation of TeV $\gamma$-rays inside 
compact massive binary systems appeared at the early 80's after first reports on 
possible observation of such $\gamma$-rays (see e.g. Weekes~1988).
The optical depths for $\gamma$-rays in the radiation field of the accretion disk
around the compact object inside the binary system have been calculated by e.g.
Carraminana~1992 and Bednarek~(1993) and in the radiation field of the massive star
by e.g. Protheroe \& Stanev~1987 and Moskalenko, Karaku\l a \& Tkaczyk~1993).
In the case of very compact binaries, the TeV $\gamma$-rays injected close to the 
surface of the massive star initiate the Inverse Compton $e^\pm$ pair cascade.
The conditions for which the cascade processes should become important
have been considered by Bednarek (1997, Sect. 2) and Dubus (2005, Sect 6).
In general, the product of the massive star luminosity and the square of its 
surface temperature should be much greater than $\sim 10^{45}$ erg s$^{-1}$ K$^2$  
(see Eg. 1 and Fig. 3 in Bednarek 1997).
The $\gamma$-ray spectra which emerge toward the observer from 
such compact binary systems strongly depend on the phase of the
injection place of the primary relativistic particles in respect to the observer.
At some phases significant TeV $\gamma$-ray fluxes are expected but at others
only photons with energies extending to a few tens GeV are able to escape (see detailed 
calculations of such anisotropic cascades and their application to specific sources,
e.g. Cen X-3 and Cyg X-3, in        
Bednarek~1997 (B97), Bednarek~2000 (B00), and Sierpowska \& Bednarek~2005 (SB05)).
Recently, optical depths of TeV $\gamma$-rays have been calculated for 
other TeV $\gamma$-ray sources, LS 5039, PSR 1259-63 and LSI +61$^{\rm o}$ 303, 
without taking into considerations the effects of the IC $e^\pm$ pair cascading 
(B\"ottcher \& Dermer~2005 and Dubus~2005).

Note that only calculations by Bednarek and collaborations (B97, B00, SB05) and 
Dubus~(2005) take into account dimensions of the massive star. Therefore, they
can be applied to very compact binaries in which injection
distance of the TeV $\gamma$-rays from the center of massive stars is be comparable 
to their radii, e.g. at the periastron passage of the compact objects
in LS 5039 and LSI +61$^{\rm o}$ 303, or in the case of propagation of $\gamma$-rays 
injected at larger distances but passing close to the surface of the massive stars.
The comparison of the exact calculations with the approximate ones, which neglect
dimensions of the massive stars, can be found in Dubus~(2005).

In this paper we concentrate on details of the propagation of high energy
$\gamma$-rays injected close to the surface of the massive star taking into account
the effects of cascades initiated through the Inverse Compton and $e^\pm$ pair 
production processes. We calculate the $\gamma$-ray spectra emerging
to the observer from such anisotropic cascades for two compact massive 
binaries LS 5039 (already reported in GeV and TeV $\gamma$-rays) and 
LSI +61$^{\rm o}$ 303 (reported only at GeV energies). 
Specific cases with the injection of primary $\gamma$-rays (e.g. by relativistic 
hadrons) or electrons (accelerated in the jet or the shock) 
with simple power law spectra and spectral indexes equal to 2
are considered in order to better understand the basic features of such 
anisotropic cascade processes.

\begin{figure}
\vskip 4.2truecm
\includegraphics{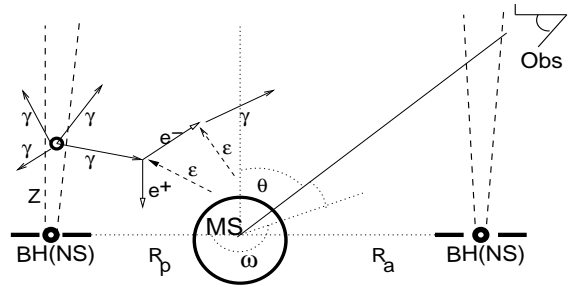}
\caption{Schematic picture of a compact binary system composed of a massive star
and a compact object (a black hole or an neutron star) on an orbit
around the massive OB star. The observer is located at the inclination angle
$\theta$ and the azimuthal angle $\omega$. The matter accreting onto a compact object 
from the massive star creates an accretion disk. Particles (electrons or protons) are 
accelerated  inside the jet launched from the inner part of an accretion disk.
Primary electrons and/or $\gamma$-rays, 
injected at the distance $z$ from the base of the jet, initiate
an anisotropic inverse Compton $e^\pm$ pair cascade 
in the radiation field of the massive star. A part of the primary $\gamma$-rays and 
secondary cascade $\gamma$-rays escape from the binary system toward the observer.}
\label{fig1}
\end{figure}
\section{The massive binary systems}

Both binary systems considered in this paper belong
to the class of the non-thermal radio high mass X-ray binaries showing evidences of 
collimated relativistic outflows. They are called microquasars due to their supposed 
similarities to quasars which show very narrow jets moving with relativistic speeds. 
In binary systems jets are launched from the inner parts of accretion disks
around compact objects (a neutron stars or a solar mass black holes).

Here we consider a simple scenario in which jets are launched along the disk axis. 
The surface of the disk is in the plane of the binary system, i.e. the jet direction 
is perpendicular to the plane of the binary system. 
The schematic picture of such binary system is drawn in 
Fig. \ref{fig1}. Relativistic particles, 
injected in the jet at the distance $z$ from its base, produce $\gamma$-rays.
If the injection site of the $\gamma$-rays is relatively close to the massive star,
$\gamma$-rays interact with the stellar radiation initiating the 
IC $e^\pm$ pair cascade. The {\it efficiency} of such cascades depend strongly on the 
parameters of the binary systems LS 5039 and LSI +61$^{\rm o}$ 303.
These two binaries differ in basic parameters, e.g their orbital periods are 
equal to 3.9 days and 26.5 days. However both of them provide conditions 
in which cascading effects have to be taken into account when considering $\gamma$-ray
production processes.

\subsection{The binary system LS 5039}

This binary system shows relativistic radio jets on miliarcsecond scales, with the 
speed of $v\sim 0.3c$ (Paredes et al. 2000). It has been also suggested to be a
counterpart of the EGRET source 3EG J1824-1514 localized at $\sim 0.5^{\rm o}$
(Paredes et al. 2005). This source has relatively flat spectrum above 100 MeV,
spectral index $<2$,
and the $\gamma$-ray luminosity $\sim 4\times 10^{35}$ erg s$^{-1}$.
Moreover, the position of LS 5039 is consistent at the $3\sigma$
level with recently detected TeV source HESS J1826-148 (Aharonian et al. 2005). 
The spectrum above 250 GeV is also flat with the photon index $2.12\pm 0.15$,  
although the luminosity is only $\sim 10^{33}$ erg s$^{-1}$, about two orders of 
magnitude less than at GeV
energies. Recent analysis of the TeV $\gamma$-ray light curve by Casares et al.~(2005b),
using new orbital parameters, show possible flux variations of a factor $\sim 3$
with the maximum around the phase $\sim 0.9$.

\begin{figure*}
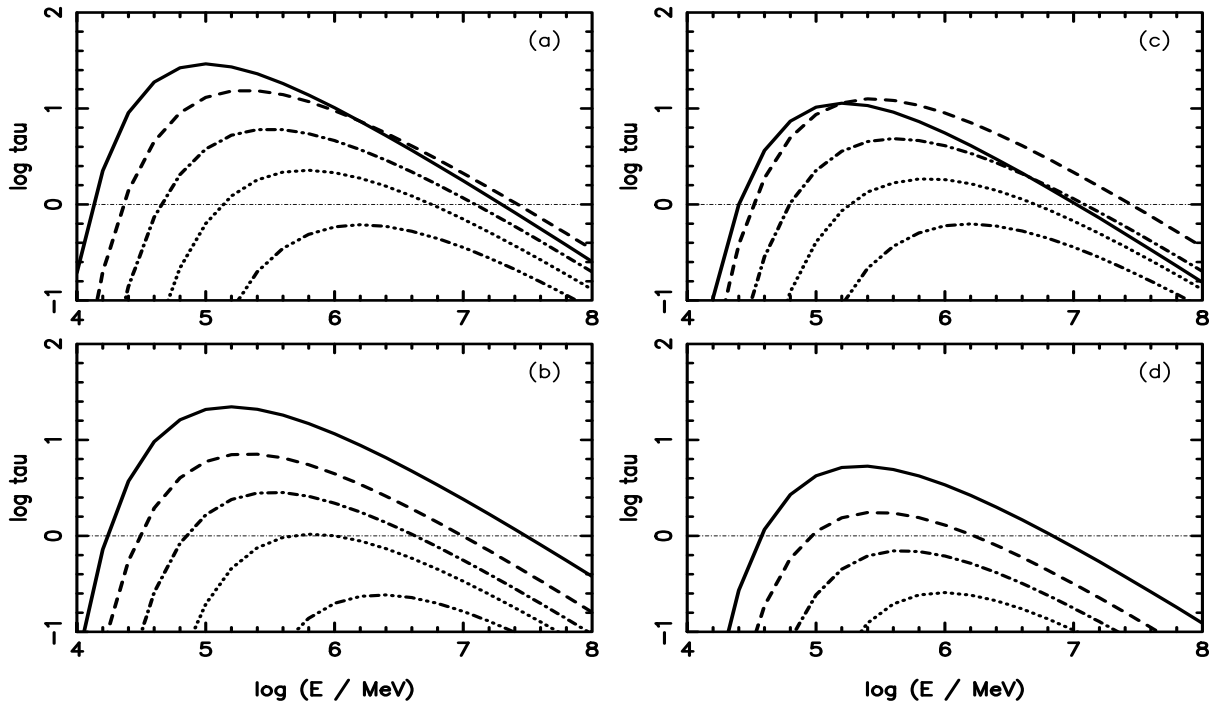

\vskip 9.5truecm
\includegraphics{taugg22.eps}
\includegraphics{taugg45.eps}
\includegraphics{taugg15.eps}
\includegraphics{taugg9.eps}
\caption{The optical depths for $\gamma$-rays (as a function of their energy) 
on $e^\pm$ pair production in collisions with stellar photons.
$\gamma$-rays are injected at two distances from the massive star in LS 5039
($2.2{\rm r}_\star$ (a) and $4.5{\rm r}_\star$ (b)) and in LSI +61$^{\rm o}$ 303 
($1.5{\rm r}_\star$ (c) and $9.15{\rm r}_\star$ (d)), 
corresponding to the periastron and the apastron passages, respectively. 
Specific curves show the optical depths for the injection angles of $\gamma$-rays 
measured from the direction defined by the centers of the stars:
$\alpha = 150^{\rm o}$ (full curve, direction toward the 
massive star), $120^{\rm o}$ (dashed), $90^{\rm o}$ (dotted), 
$60^{\rm o}$ (dot-dashed), and $30^{\rm o}$ (dot-dot-dot-dashed).
The optical depths are calculated up to the infinity accept two cases (marked by full
curves in (a) and (c)) which are calculated to the surface of the massive star.}
\label{fig2}
\end{figure*}

The basic parameters of the binary system LS 5039 has been recently
reported by Casares et al.~(2005b):
the semimajor axis, $a = 3.4 r_\star$, ellipticity $e = 0.35\pm 0.04$,
the azimuthal angle of the observer in respect to the periastron passage,
$\omega = 225^{\rm o}$, radius of the massive star, 
$r_\star = 9.3^{+0.7}_{-0.6}$ R$_\odot$, and its surface temperature, 
$T_{\rm s} = 3.9\times 10^4$ K.
For these parameters the distance of the compact object from the 
massive star chances in the range: $r_{\rm p} = 2.2 r_\star$ at the periastron up to
$r_{\rm a} = 4.5 r_\star$ at the apastron.
The estimated inclination angle of the binary system toward the observer depends on
the mass of the compact object. It is estimated on $\theta= 24.9\pm 2.8^{\rm o}$ for the 
case of the black hole with the mass $3.7M_\odot$ and on $\sim 60^{\rm o}$ for the neutron star
(Casares et al.~2005b). We present the results of numerical calculations for both 
inclination angles in the case of LS 5039.

\subsection{The binary system LSI +61$^{\rm o}$ 303}

LSI +61$^{\rm o}$ 303 has been also observed as a non-spherical radio source, 
which structure was interpreted as due to relativistic radio jets, the speed 
of $\sim 0.6c$, with some hints of its precession (Massi et al. 2004). 
It has been pointed out (Gregory \& Taylor~1978) that this source is connected with 
the COS B $\gamma$-ray source CG135+01 (Hermsen et al. 1977). 
CG135+01 has been also detected by EGRET 
($>$100 MeV); the source 2EG J0241+6119 has a hard spectrum with photon spectral index 
$2.05\pm 0.06$ (Kniffen et al. 1997), and by COMPTEL in the energy range $\sim$0.75 
to 30 MeV, spectral index $1.95^{+0.2}_{-0.3}$ (van Dijk et al. 1996). 
The analysis of different EGRET observations
shows evidences of variability (Tavani et al. 1998, confirmed by Wallace 
et al. 2000) with probable modulation with the orbital period of  LSI +61$^{\rm o}$ 303
and the maximum emission near the periastron passage (Massi 2004).
The Whipple group (Hall et al.~2003) puts an upper limit on the TeV flux from 
this source, $\sim 10^{-11}$ cm$^{-2}$ s$^{-1}$ above 500 GeV ($< 1.3\times 10^{32}$ 
erg s$^{-1}$), which is clearly below an extrapolation of the EGRET spectrum.

The basic parameters of the binary system LSI +61$^{\rm o}$ 303 has been recently
reported by Casares et al.~(2005a):
the semimajor axis $a = 5.3 r_\star$, ellipticity $e = 0.72$,
the inclination of the binary system toward the observer is not well constrained by 
the observations (Casares et al. 2005): $25^{\rm o} < i < 60^{\rm o}$ for a neutron 
star and $\theta< 25^{\rm o}$ for a black hole. We apply the value of 
$\theta= 30^{\rm o}$. 
The azimuthal angle of the observer in respect to the periastron passage is
$\omega = 70^{\rm o}$, radius of the massive star 
$r_\star = 13.4$ R$_\odot$, and its surface temperature $T_{\rm s} = 2.8\times 10^4$ K.
For these parameters the distance of the compact object from the 
massive star changes in the range: $r_{\rm p} = 1.5r_\star$ at the periastron up to
$r_{\rm a} = 9.15r_\star$ at the apastron.
We apply these parameters in our further calculations.

\section{Optical depths for $\gamma$-rays}

The optical depths for $\gamma$-rays (for the process 
$\gamma + \gamma\rightarrow e^+e^-$) in the radiation field of the massive star,
determined by the radius of the star and its surface temperature,
are calculated in the general case, i.e. for an arbitrary place of injection of
$\gamma$-ray photons with arbitrary energies and angles of propagation
(see also B97, B00, SB05). This approach allows us to calculate the optical depths 
even for the primary $\gamma$-rays
injected at the surface of the massive star. Therefore, it can be applied for 
studies of the cascade processes initiated by $\gamma$-ray photons since
secondary $\gamma$-rays can appear, in principle, everywhere inside the binary
system. 
In fact, the optical depths for the $\gamma$-rays propagating in the thermal
radiation of these massive stars can be obtained by simple 
re-scaling of the earlier calculations for the massive star
in Cen X-3 (shown e.g. in Bednarek 2000, Fig.~2), since they are 
proportional to the fourth power of the surface temperature and the square of the radius
of the massive star. Also a shift in energy proportional to the surface temperature is necessary. 
Note, that in the case of cascade processes discussed in this paper, primary particles
propagating at specific direction can contribute to the final
$\gamma$-ray spectra escaping at other directions. Therefore, 
for easier analysis of the obtained results,
we show here the optical depths for $\gamma$-rays injected at the distance of the 
periastron and the apastron passages of the compact object
around the massive stars in LS 5039 (Fig.~\ref{fig2}a,b) and LSI +61$^{\rm o}$ 303
(Fig.~\ref{fig2}c,d), as a function of the photon energy and their arbitrary 
injection angles, 
$\alpha$, measured from direction defined by the centers of the companion stars,
toward the massive star. 

As expected, the optical depths strongly depend on the injection parameters 
(photon energy, angles of propagation) and the parameters of the massive star.
The optical depths reaches the maximum corresponding to the peak in the black body
spectrum of soft photons. The maximum optical depth shifts to larger photon 
energies with decreasing angle $\alpha$ (provided that $\alpha$ is smaller than the 
angle, $\beta$, intercepted by the massive star observed from the distance of 
the injection place). The optical depths are lower for $\alpha > \pi - \beta$, 
since in this case $\gamma$-rays propagate only 
to the surface of the massive star (see full thick curves in Figs.~\ref{fig2}a and c).
$\gamma$-rays with energies $\sim 1$ TeV, injected at the periastron distance in 
LS 5039, $2.2r_\star$, are absorbed 
($\tau > 1$) for most of the propagation directions, except a small  
cone with the angular extend of $\sim 40^{\rm o}$. At the apastron distance, 
$4.5r_\star$, the escape cone increases to $\sim 60^{\rm o}$.
Only within such small cones, $\gamma$-rays with $\sim 1$ TeV energy have high chance to escape
without absorption.  The optical depth for $\gamma$-rays
in LSI +61$^{\rm o}$ 303
are typically lower due to significantly lower surface temperature of the
massive star. At the periastron distance, $1.5r_\star$, the escape cone for 
$\sim 1$ TeV $\gamma$-rays is also $\sim 40^{\rm o}$, very similar to the periastron 
distance in LS 5039, due to the closer location of the injection place.
However, at the apastron distance, $9.15r_\star$, most of $\sim 1$ TeV 
$\gamma$-rays escape from the binary system without absorption, except $\gamma$-rays 
moving toward the massive star within the cone with the angle $\sim 80^{\rm o}$. 

In LS 5039, the angles between the observer (at the inclination angle $25^{\rm o}$) 
and the direction defined by the stars are $\sim 110^{\rm o}$ for 
the periastron passage and $\sim 70^{\rm o}$ for the apastron passage.
For such geometry, the optical depths towards the observer
are larger than unity for $\gamma$-rays with energies in the range 
$\sim 0.03 - 20$ TeV, for the periastron passage, and $\sim 0.2 - 2$ TeV, 
for the apastron 
(see Fig.~\ref{fig2}a,b). These calculations of the optical depths are 
generally consistent with that obtained by Dubus (2005).

In LSI +61$^{\rm o}$ 303, the observer is located at the angle of $\sim 80^{\rm o}$ at 
the periastron passage and at $\sim 100^{\rm o}$ at the apastron passage.
In this case the optical depths
are larger than unity for $\gamma$-rays with energies in the range $\sim 0.1 - 10$ TeV 
for the periastron passage, but always lower than unity for the apastron, provided 
that they are injected toward the observer (see Fig.~\ref{fig2}c,d).

\section{The cascade $\gamma$-ray spectra}

Since the optical depths for high energy $\gamma$-rays in the radiation fields
of the massive stars in binary systems LS 5039 and LSI +61$^{\rm o}$ 303 are large,
the total $\gamma$-ray spectra, emerging from the binary systems toward the observer,
are determined by the primary $\gamma$-ray spectra (if such are produced e.g. by hadrons
accelerated in the jet) and by the $\gamma$-ray spectra produced in
the IC $e^\pm$ pair cascades occurring in the radiation of the massive star. 
However, since the injection place of primary $\gamma$-rays or primary electrons
is different than the center of the massive star (the source of isotropic soft 
radiation), the IC $e\pm$ pair cascade occurs in fact in the anisotropic radiation 
field. IC $e^\pm$ pair cascades of such type have been already 
considered by us under two extreme assumptions. In the first approximation, 
local isotropization of the secondary cascade $e^\pm$ pairs by the random component
of the magnetic field inside the stellar wind and the jet is assumed (see B97, B00). 
In the second approach, we follow the paths of secondary cascade $e^\pm$ pairs
in the magnetic field which structure is described by a specific model (see SB05).
In the present work we apply the first approach. 
Note, that the IC $e^\pm$ pair cascade considered in this work 
does not take into account the synchrotron losses of $e^\pm$ pairs and assumes that 
secondary $e^\pm$ pairs radiate locally secondary $\gamma$-rays, i.e. at their production site.
The conditions for such type of cascade
are determined in Bednarek (B97, see Sect. 2). For completeness and due to some 
different details of the scenario considered in this paper, we discuss below
some important conditions for IC $e^\pm$ pair cascades occurring in the radiation
field of the massive star.

Leptons injected into the radiation of the massive star (primary electrons and secondary 
$e^\pm$ pairs from the cascade)
are immersed in the stellar wind and/or in the plasma outflow along the jet.
Therefore, efficiency of the ICS process is determined by the 
relative importance of the IC cooling time scale in respect to the characteristic 
escape time scale, determined by the advection time scale of $e^\pm$
pairs with the stellar and/or jet outflows. 
The IC cooling time of leptons, the Thomson regime, in the radiation of the massive star 
can be estimated from,
\begin{eqnarray}
\tau_{\rm IC} = m_{\rm e}\gamma_{\rm e}/{\dot {\rm P}}_{\rm IC}^{\rm T}
\approx 4\times 10^5r^2/(\gamma_{\rm e}T_{\rm 4}^4)~~~{\rm s},
\label{eq1}
\end{eqnarray}
\noindent
where $m_{\rm e}$ and $\gamma_{\rm e}$ are the rest mass and the Lorentz factor of leptons,
${\dot {\rm P}}_{\rm IC}^{\rm T} = (4/3) c \sigma_{\rm T}\gamma_{\rm e}^2$ is the energy 
loss rate on ICS of leptons, $c$ is the velocity of light, 
$\sigma_{\rm T}$ is the Thomson cross section, 
$U_{\rm rad} = \sigma_{\rm SB}T_{\rm s}^4/r^2$, $\sigma_{\rm SB}$ is the Stefan-Boltzmann 
constant, $T_{\rm s} = 10^4T_{\rm 4}$ K and $R = r r_\star$ are the surface 
temperature of the massive star and the distance from the center of the massive star 
(in units of its radius).
The characteristic escape time scale of leptons from their
creation (acceleration) place, identified with their advection time scale
with the jet or the wind plasma flow, is 
\begin{eqnarray}
\tau_{\rm esc} = R/V \approx 33r_{\star,12}r/v~~~{\rm s},
\label{eq2}
\end{eqnarray}
\noindent
where $v = V/c$ is the velocity of the jet (or the stellar wind). It is assumed that 
typical radii of the massive stars are of the order of 
$r_\star = 10^{12}r_{\star,12}$ cm.
By comparing Eq.~1 and Eq.~2, we estimate the minimum 
Lorentz factor of leptons above which they cool locally,
\begin{eqnarray}
\gamma_{\rm e} > \gamma_{\rm min}\approx 1.2\times 10^4 r v/(T_{\rm 4}^4r_{\star,12}).
\label{eq3}
\end{eqnarray}
\noindent
For example, in the case of the massive star in LS 5039, $T_{\rm 4} = 3.9$, 
$r_{\star,12} = 1$, and $v = 0.5$, leptons with the Lorentz factors above, 
$\gamma_{\rm min}\approx 25r$, cool locally before escaping from the binary system.

Leptons with Lorentz factors, $\gamma_{\rm min}$, produce $\gamma$-ray photons 
with energies, 
\begin{eqnarray}
E_\gamma\approx  (4/3)\epsilon\gamma_{\rm min}^2\approx 10^3 r^2 
v^2/(T_{\rm 4}^7r_{\star,12}^2)
~~~{\rm MeV},
\label{eq4}
\end{eqnarray}
\noindent
where typical energies of photons coming from the massive stars are 
$\epsilon = 3k_{\rm B} T_{\rm s}\approx 2.6\times 10^{-6}T_{\rm 4}$ MeV.
We conclude that spectra of $\gamma$-ray photons produced in the 
cascade (assuming local cooling of leptons) are correct above energies
given by Eq.~\ref{eq4}. For the parameters of the massive stars in LS 5039 and 
LSI +61$^{\rm o}$ 303, escaping $\gamma$-ray spectra are correct above
$\sim 10^3$ MeV, provided that production of $\gamma$-rays occur inside the jet and 
within $r\sim 30r_\star$ from the center of the massive star.
For the stellar wind region, the limits on the 
low energy cut-offs in the cascade $\gamma$-ray spectra are about two orders of magnitude
lower due to much lower stellar wind velocities in respect to the plasma velocity
in the jet.

However, for the Lorentz factors above $\sim 10^5$ scattering of soft photons 
occurs in the Klein-Nishina (KN) regime. We approximate the 
IC cooling time of leptons in the KN regime by,
\begin{eqnarray}
\tau_{\rm IC}^{\rm KN} = m_{\rm e}\gamma_{\rm e}/{\dot {\rm P}^{\rm T}}_{\rm IC}(\gamma_{\rm KN/T})
\approx 8\times 10^{-6}r^2\gamma_{\rm e}/T_{\rm 4}^2~~~{\rm s},
\label{eq1}
\end{eqnarray}
\noindent
where $\gamma_{\rm KN/T}\approx m_{\rm e}c^2/\epsilon\approx 2\times 10^5/T_4$. 
By comparing the escape time scale, $\tau_{\rm esc}$, with the IC cooling time scale in 
the KN regime $\tau_{\rm IC}^{\rm KN}$,
we obtain the upper limit on the Lorentz factor of leptons which are able to 
cool locally 
\begin{eqnarray}
\gamma_{\rm e} < \gamma_{\rm max}\approx 4\times 10^6 r_{\star,12}T_{\rm 4}^2/(r v).
\label{eq3}
\end{eqnarray}
\noindent
Therefore, we conclude that leptons with energies $\sim 10$ TeV should be able to cool locally within a few
stellar radii from the surface of the massive stars in LS 5039 and LSI +61$^{\rm o}$ 303.

In the calculations shown below we assume that synchrotron energy losses of leptons 
can be neglected in respect to the IC losses. This is correct provided that
the surface magnetic fields of the massive stars fulfill the condition,
\begin{eqnarray}
B_{\rm s} < 40T_{\rm 4}^2~~~{\rm G},
\label{eq5}
\end{eqnarray}
\noindent
obtained from the comparison of the synchrotron and IC energy losses of leptons
in the Thomson regime. The simple limit given by Eq.~\ref{eq5} is valid under 
assumption on the $B\propto r^{-2}$ magnetic field dependence on the distance from
the stellar surface. It neglects the inner dipole part of the magnetic field in which 
$B\propto r^{-3}$. Therefore, real upper bound on the surface magnetic field
is a factor of 2-3 larger than given by Eq.~\ref{eq5}. 
For the massive stars in LS 5039 and LSI +61$^{\rm o}$ 303, these upper bounds 
(Eq.~\ref{eq5}) are $\sim 620$ G and $\sim 320$ G.
In the Klein-Nishina regime, i.e. for $\gamma_{\rm e}\gg 10^5$, the limit on the 
magnetic field is more restrictive due to the decrease of the cross section. It has 
been discussed in details by Bednarek (B97, see Fig.~4 in that paper). 

Under conditions specified above (neglected synchrotron losses of leptons and their local 
isotropization), we study the features of the $\gamma$-ray 
spectra emerging from the binary system, due to the propagation effects in 
the radiation of 
their massive companions, assuming that primary particles, i.e. electrons or 
$\gamma$-rays, are isotropically 
injected somewhere 
along the jet. We consider the primary particles with the power law spectra
and spectral index equal to 2 due to the equally distributed power per decade.
The high energy cut-off in these primary spectra is assumed at 10 TeV 
(to be consistent with the observations of $\gamma$-ray spectrum from LS 5039 up to 
$\sim 4$ TeV, Aharonian et al.~(2005)). These initial spectra of particles 
are normalized to unity.
The spectral index equal to 2 is also motivated by a relatively flat GeV 
$\gamma$-ray spectra observed from these two sources. 
The jets in microquasars move with much lower velocities (Doppler factor $D\sim 1$) 
than these ones observed in  active galactic nuclei. Therefore, the effects of 
relativistic beaming can be neglected in the first approximation.

\subsection{LS 5039}

\begin{figure*}
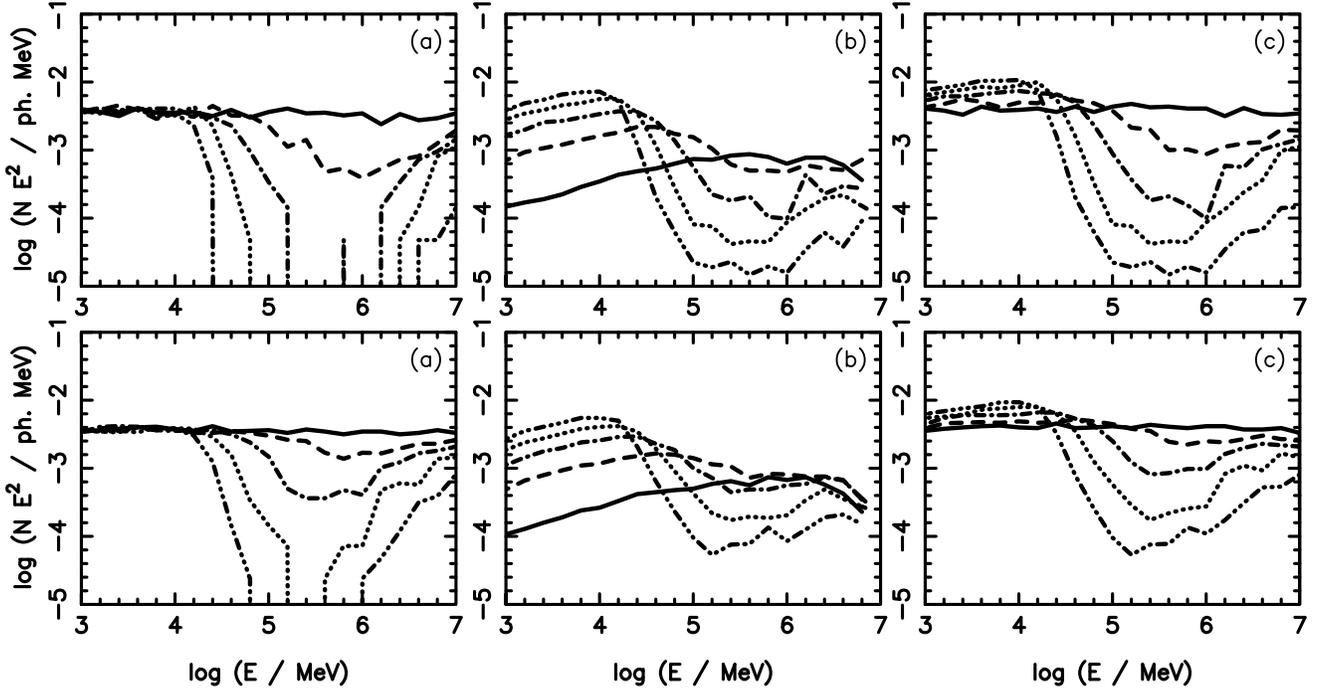

\vskip 9.5truecm
\includegraphics{spec22suma.eps}
\includegraphics{spec22wtor.eps}
\includegraphics{spec22pier.eps}
\includegraphics{spec45suma.eps}
\includegraphics{spec45wtor.eps}
\includegraphics{spec45pier.eps}
\caption{Differential $\gamma$-ray spectra (multiplied by the square of 
photon energy) escaping from the binary system at a  
specific range of the cosine angles $\alpha$, measured in respect to 
the direction defined by the injection place and the center of the massive star. 
The range of $\Delta\cos\alpha$ (with the width 0.1) are centered on 
0.95 (full curve), 0.55 (dashed), 0.15 (dot-dashed), 
-0.25 (dotted), and -0.65 (dot-dot-dot-dashed).
$\gamma$-rays are produced in the cascade initiated by
primary $\gamma$-rays with the power law spectrum and spectral index 2 which are
injected isotropically close to the base of the jet ($z\approx 0\ll r_\star$, but 
sufficiently far away from the accretion disk), and at two distances from the 
massive star $2.2r_\star$ 
(upper figures, corresponding to the periastron passage of the compact object) and 
$4.5r_\star$ (bottom figures, corresponding to the apastron passage).
(a) Spectra of primary $\gamma$-rays escaping without interaction.
(b) Spectra of secondary $\gamma$-rays produced in the IC cascade.
(c) Total spectra of $\gamma$-rays escaping from the binary system (the sum of spectra 
shown in (a) and (b)).}
\label{fig3}
\end{figure*}
\begin{figure*}
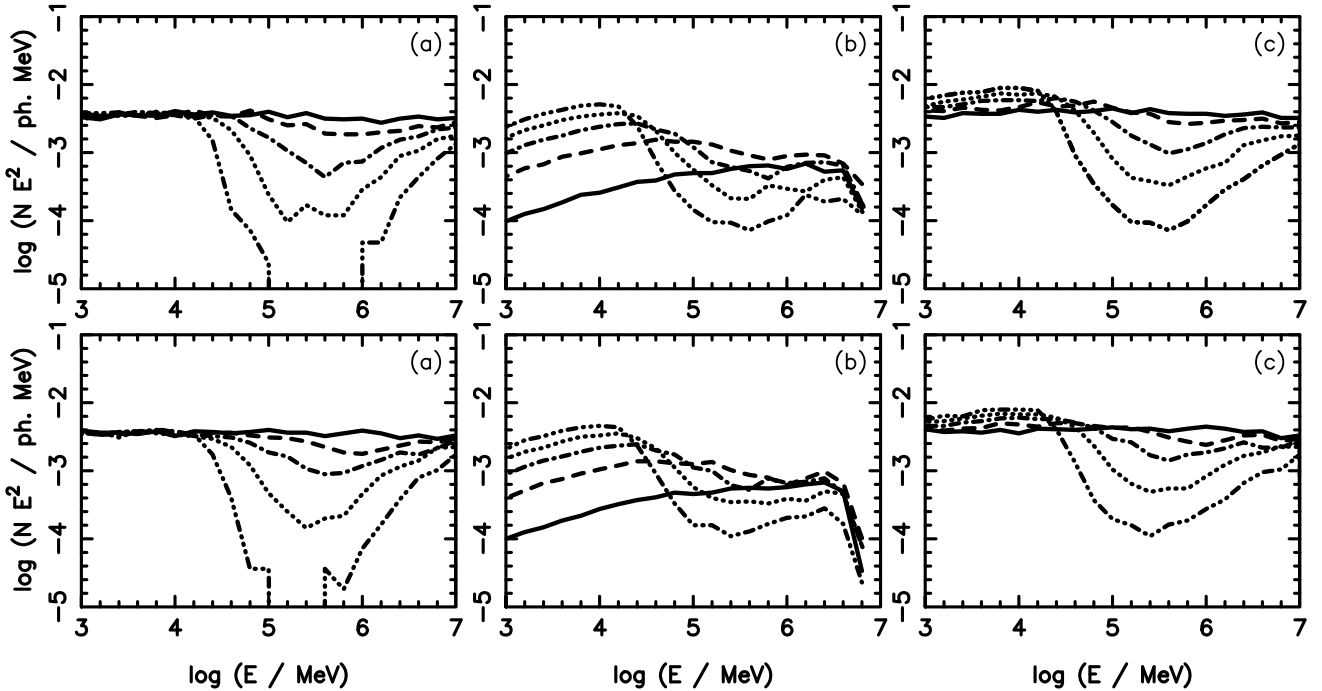

\vskip 9.5truecm
\includegraphics{spec22sumaz5.eps}
\includegraphics{spec22wtorz5.eps}
\includegraphics{spec22pierz5.eps}
\includegraphics{spec45sumaz5.eps}
\includegraphics{spec45wtorz5.eps}
\includegraphics{spec45pierz5.eps}
\caption{As in Fig.~\ref{fig3} but for the injection distance along the jet 
$z = 5r_\star$. The distance of the injection place from the massive star is
$\sim 5.5r_\star$ at the periastron (upper figures) and $\sim 6.7r_\star$ at the 
apastron (bottom figures).}
\label{fig4}
\end{figure*}
\begin{figure*}
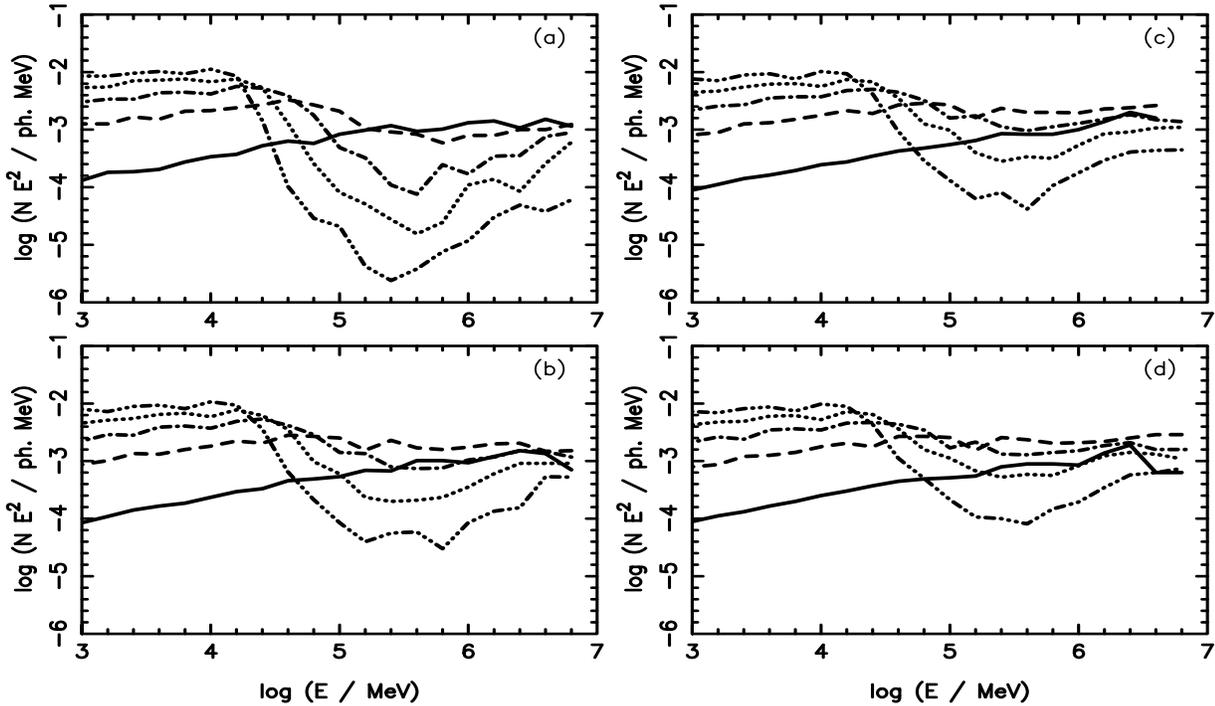

\vskip 9.5truecm
\includegraphics{specelec22.eps}
\includegraphics{specelec45.eps}
\includegraphics{specelec22z5.eps}
\includegraphics{specelec45z5.eps}
\caption{Total $\gamma$-ray spectra from the cascade initiated by isotropically
injected primary leptons, with the power law spectrum and spectral index 2,
at the range of cosine of the observation angles $\cos\alpha$, 
with the width $\Delta\cos\alpha=0.1$. The spectra are shown for $\cos\alpha$ centered 
on 0.95 (full curve), 0.55 (dashed), 0.15 (dot-dashed), 
-0.25 (dotted), and -0.65 (dot-dot-dot-dashed),
for two distances of the injection place from the massive star $2.2r_\star$ 
(figures (a) and (c)) and $4.5r_\star$ (figures (b) and (d)), corresponding to the 
periastron and the apastron of the injection place, respectively. 
The spectra are shown for the injection place at the base of the jet
$z\approx 0\ll r_\star$ ((a) and (b)) and at the distance of z=5r$_\star$ 
from the base of the jet ((c) and (d)).}
\label{fig5}
\end{figure*}

Let's at first consider the case of isotropic injection of primary electrons 
at the base of the jet, i.e. $z\approx 0$.
In fact, $\gamma$-rays have to be injected at some distance from the
accretion disk at which the disk radiation can be neglected in respect to the 
stellar radiation.  
The distance from the base of the jet at which the above condition can be fulfilled 
is estimated by comparing the 
energy density of the disk radiation with the energy density of the stellar radiation.
These radiation fields are defined by the surface temperature of the massive star 
and the maximum temperature on the surface of the disk (at the disk inner radius, 
$T_{\rm in}$). We approximate the accretion disk radiation by the model of Shakura \&
Sunyaev (1973) in which the disk radiation from its surface can be approximated by
the black body radiation with some temperature gradient. 
The following condition has to be approximately fulfilled,
\begin{eqnarray}
T_{\rm s}^4/r^2\approx T_{\rm in}^4(r_{\rm in}/z)^2, 
\label{eq6}
\end{eqnarray}
\noindent
where $r_{\rm in}$ is the disk inner radius. From this condition, we estimate that 
above,
\begin{eqnarray}
z\approx r_{\rm in}r(T_{\rm in}/T_{\rm s})^2, 
\label{eq7}
\end{eqnarray}
\noindent
the massive star radiation dominates over the accretion disk radiation. 
For typical parameters of the considered binary systems, $T_{\rm s} = 3\times 10^4$ K,
$r = 10$, the accretion disk inner radius $r_{\rm in} = 10^7$ cm, and 
$T_{\rm in} = 10^6$ K
(limited by the condition that the disk thermal luminosity has to be lower 
than the observed X-ray luminosity from these binary systems, $< 10^{35}$ erg s$^{-1}$), 
we estimate that above $z\sim 10^{11}$ cm ($\sim 0.1r_\star$) from the base 
of the jet, stellar radiation dominates over accretion disk radiation.  
This condition is valid in the Thomson regime. However for the IC scattering process
occurring in the Klein-Nishina regime, stellar radiation starts to dominate even at 
lower distances from the inner part of the disk, due to significantly larger 
temperature
in the inner disk than on the stellar surface. These same arguments concern also
possible disk corona which energy density can be comparable to the energy density of 
radiation from the surface of the disk but the characteristic temperature 
(average photon energies) are larger.
The conditions for absorption of $\gamma$-ray photons in the radiation of the massive
star are more favorable in respect to the radiation of the accretion disk 
due to larger angular extend
of the stellar disk in respect to the inner part of the accretion disk. 
In conclusion, we consider the injection of the primary $\gamma$-rays and electrons 
at the base of the jet
assuming that it occurs at the distance, $z > 0.1r_\star\approx 0$, which is
very close to the base of the jet in respect to dimensions of the massive star.

Although primary $\gamma$-rays and soft photons from the massive star
are injected isotropically, their injection places are located at different 
parts of the binary system (the jet or the compact object for primary 
$\gamma$-rays and the massive star for soft photons). Therefore,
in fact primary $\gamma$-rays develop an IC e$^\pm$ pair cascade  in the
non-isotropic radiation field. The spectra, which emerge from the binary system,    
depend on the location of the observer in respect to direction defined
by the location of the injection place and the center of the massive star 
(the angle $\alpha$ measured from the outward direction defined by the injection place
and the center of the massive star).
We calculate the angle dependent spectra of $\gamma$-rays escaping from the binary system,
by applying the Monte Carlo method (details are described in Bednarek~B00).
Such method allows us to include the re-distribution of directions of secondary cascade
$\gamma$-rays in respect to directions of their parent primary $\gamma$-rays, which is due 
to the isotropization of secondary cascade $e^\pm$ pairs and their preferable
head on interactions with soft photons arriving from specific direction on the sky
(i.e. inside the stellar disk limb).
The parameters of $\gamma$-ray photons produced in the cascade
(their energies and the escape angles 
$\alpha$) are sorted at a specific range of the cosine angles, $\cos\alpha$, with 
the width of $\Delta\cos\alpha = 0.1$. 

We show the spectra of primary $\gamma$-rays which 
escape from the radiation field of the massive star without absorption
(Fig.~\ref{fig3}a), the spectra of $\gamma$-rays produced as a secondary cascade 
products (Fig.~\ref{fig3}b), and the sum of these
two, i.e. the total $\gamma$-rays spectra (Fig.~\ref{fig3}c), for primary 
$\gamma$-rays injected at the distance 
$z\approx 0$ along the jet, and for two locations of the 
compact object on its orbit, at the periastron distance from the massive star
($2.2r_\star$, upper figures in Fig.~\ref{fig3}) and at 
the apastron distance ($4.5r_\star$, bottom figures).
Simple absorption effects of primary $\gamma$-rays, for both distances
of the injection place from the massive star, are very strong (see Fig.~\ref{fig3}a).
Primary $\gamma$-rays with energies in the range between a few tens of GeV up to 
a few TeV are completely absorbed (the exact range depends strongly on the observation 
angles), provided that the observation angles lay within the hemisphere containing 
the massive star. However, these strong deficits of $\gamma$-rays (between $0.1 - 1$ TeV)
are partially fulfilled by the secondary cascade $\gamma$-rays (see Fig.~\ref{fig3}b). 
The secondary $\gamma$-rays contribute also to the total $\gamma$-ray spectrum below
$\sim 10$ GeV escaping in the outward directions, i.e. for $\alpha> 90^{\rm o}$ 
(Fig.~\ref{fig3}b). Total $\gamma$-ray spectra (Fig.~\ref{fig3}c)
strongly depend on the observation angle. they show strong deficit above $\sim 100$ GeV
(a dip up to two orders of magnitudes) and the excess (up to a factor of two) 
in respect to the shape of the primary power law spectrum for large angles $\alpha$.
However, they become more similar to the injected $\gamma$-ray spectrum
for small $\alpha$. As expected, $\gamma$-ray spectra 
escaping from the binary system for the injection places located at larger distances 
from the massive star (at the apastron passage), are
less modified by the cascading processes than these ones produced at the periastron
passage. The secondary $\gamma$-ray spectra produced by primary $\gamma$-rays injected 
at the apastron passage show decline above a few TeV (Fig. 3b). 
This is due to the high energy cut-off in the spectrum of primary particles at 10 TeV  
and due to the average larger interaction angles between primary $\gamma$-rays and soft 
photons from the massive star at the apastron passage.

\begin{figure*}
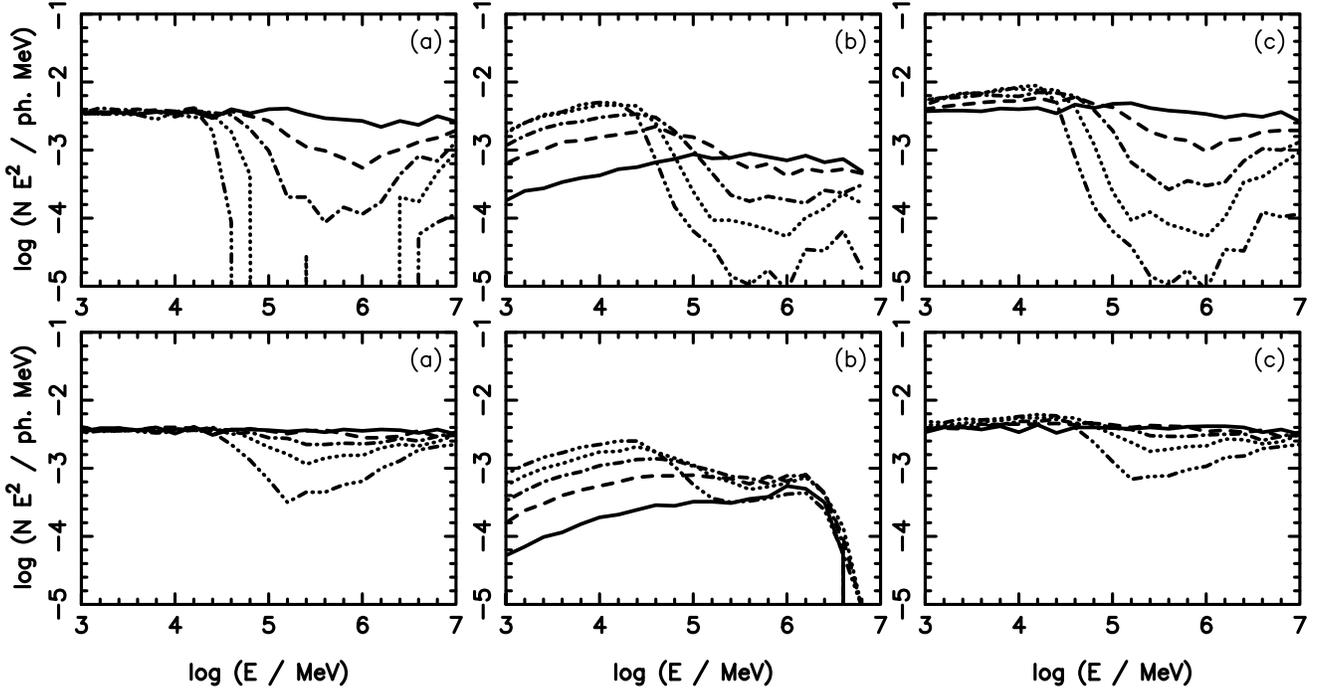

\vskip 9.5truecm
\includegraphics{spec15suma.eps}
\includegraphics{spec15wtor.eps}
\includegraphics{spec15pier.eps}
\includegraphics{spec915suma.eps}
\includegraphics{spec915wtor.eps}
\includegraphics{spec915pier.eps}
\caption{Differential $\gamma$-ray spectra escaping from the binary system at a  
specific range of the cosine angle $\alpha$ (as in Fig.~\ref{fig3}) but for the  
binary system LSI +61$^{\rm o}$ 303.
The injection place of primary $\gamma$-rays is at the distance of the periastron 
passage ($R = 1.5r_\star$, upper figures) and the apastron passage
($9.15r_\star$, bottom figures) close to the base of the jet ($z\approx 0\ll r_\star$).
The primary $\gamma$-rays spectra which escape without absorption are in (a), 
secondary cascade $\gamma$-ray spectra in (b), and the sum of (a) and (b), total 
escaping $\gamma$-ray spectra are in (c).}
\label{fig6}
\end{figure*}
\begin{figure*}
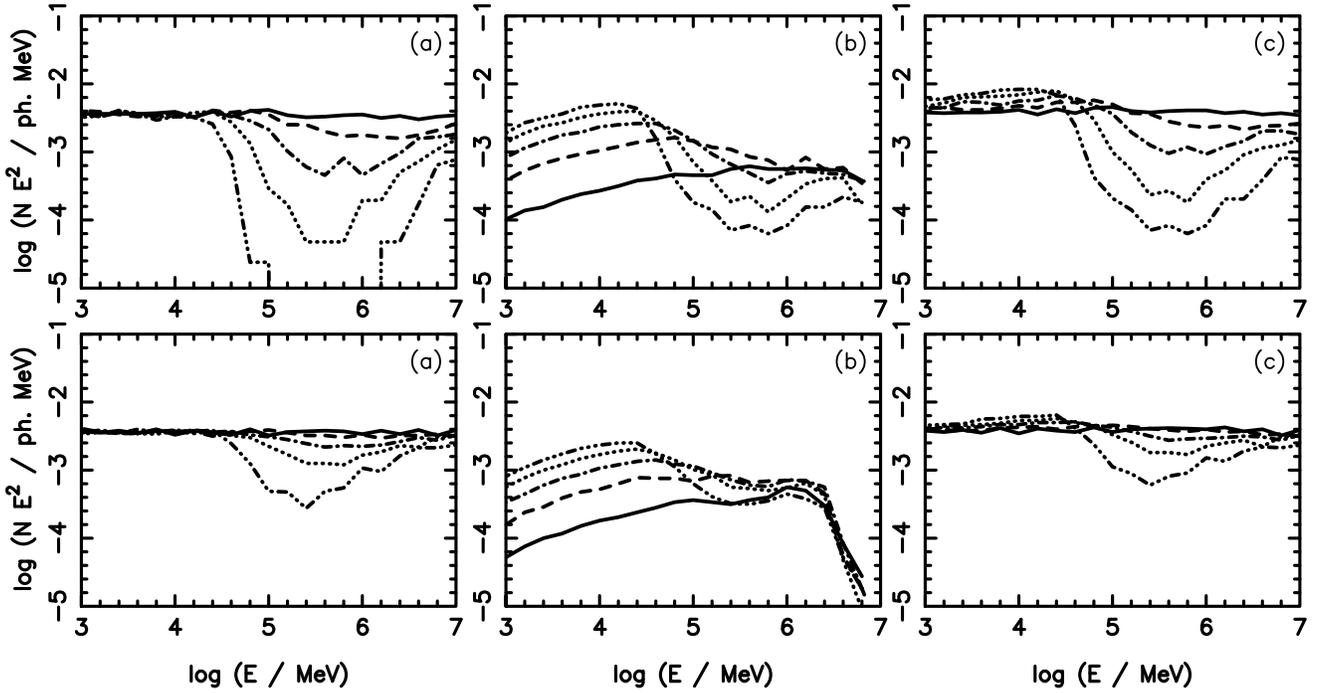

\vskip 9.5truecm
\includegraphics{spec15sumaz3.eps}
\includegraphics{spec15wtorz3.eps}
\includegraphics{spec15pierz3.eps}
\includegraphics{spec915sumaz3.eps}
\includegraphics{spec915wtorz3.eps}
\includegraphics{spec915pierz3.eps}
\caption{The $\gamma$-ray spectra produced by primary $\gamma$-rays (as in 
Fig.~\ref{fig6}) but for the injection place of primary $\gamma$-rays and the
distance $z = 3r_\star$ from the base of the jet.
The distance of the injection place from the massive star is
$\sim 3.35r_{\rm star}$ at the periastron (upper figures) and $\sim 9.6r_\star$ at the 
apastron (bottom figures).}
\label{fig7}
\end{figure*}
\begin{figure*}
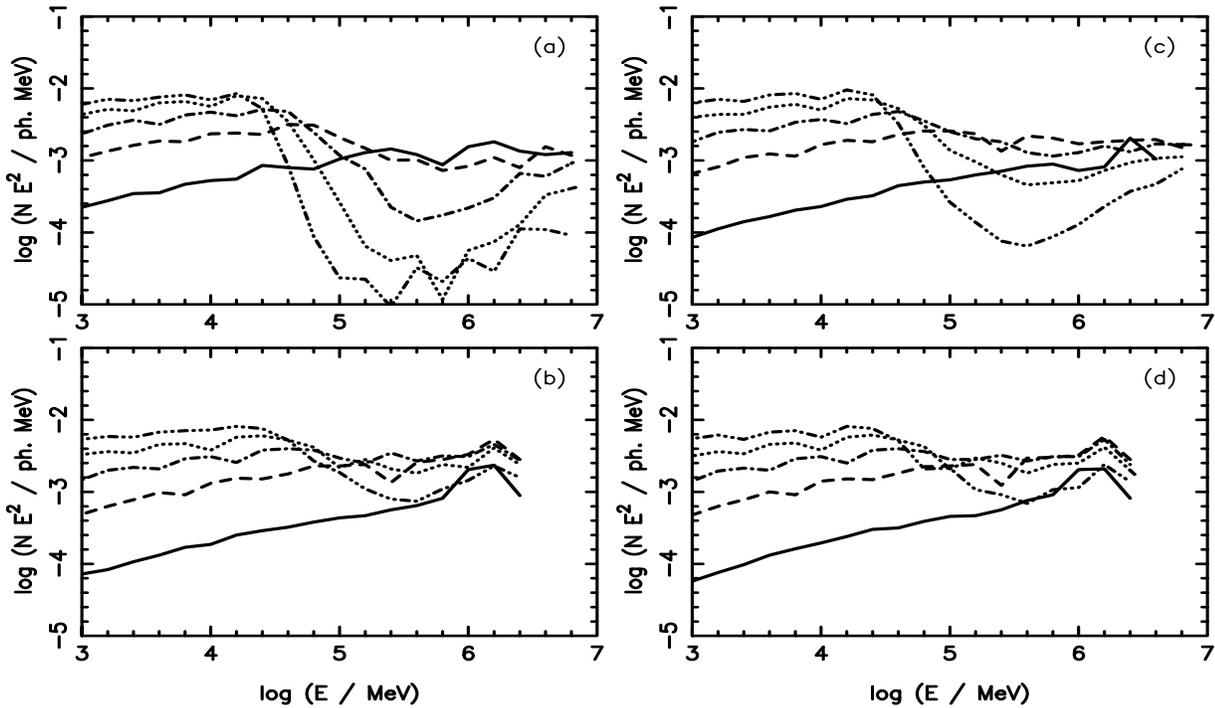

\vskip 9.5truecm
\includegraphics{specelec15.eps}
\includegraphics{specelec915.eps}
\includegraphics{specelec15z3.eps}
\includegraphics{specelec915z3.eps}
\caption{As in Fig.~\ref{fig5} but for LSI +61$^{\rm o}$ 303.
$\gamma$-ray spectra are calculated for the injection place
of primary electrons at the periastron distance (i.e. at $1.5r_\star$ - figures (a) 
and (c)) and at the apastron distance (i.e. at $9,15r_\star$ - figures (b) and (d)), 
and at the distance from the base of the jet $z = 0$ (figures (a) and (b)),
and $z = 3R_\star$ (figures (c) and (d)).}
\label{fig8}
\end{figure*}

In Fig.~\ref{fig4} we also show the $\gamma$-ray spectra produced in the cascade
for the periastron and the apastron distances assuming that the injection place is
located at the distance of $z = 5r_\star$ from the base of the jet in order to 
have impression how significant are the $\gamma$-ray spectra on the production site
in the jet. General features of these $\gamma$-ray spectra
are quite similar to the case of injection at the base of the jet ($z\approx 0$).
Therefore, if the injection of primary $\gamma$-rays occurs with similar acceleration
efficiency and spectrum along the jet within a few stellar radii from the base of the
jet, then the angular distribution of $\gamma$-rays (and their spectra) formed in 
the cascade process show quite similar 
features in relation to the direction defined by the injection place and 
the massive star. Note however, that the $\gamma$-ray spectra toward the observer 
located at fixed direction in respect to the plane of the binary may look very different
for these two injection places as we discuss in Sect. 5.

As a second scenario we consider the injection of primary electrons in the specific
region of the jet. As above, we consider two locations for the injection 
place along the jet: the base of the jet (Fig.~\ref{fig5}a,b) and the distance 
$z = 5r_\star$ from the base of the jet (Fig.~\ref{fig5}c,d). The case of isotropic 
injection of primary electrons with the power law spectrum differs 
from previously considered case of 
isotropic injection of primary $\gamma$-rays since electrons have tendency of 
more frequent production of the first generation of cascade $\gamma$-rays 
in the direction toward 
the massive star (the effect of non-isotropic radiation and kinematics of IC process).
Therefore, $\gamma$-ray spectra which escape from the binary system in this second scenario
resemble the secondary cascade $\gamma$-ray spectra produced in the case of injection of
primary $\gamma$-rays (compare Figs.~\ref{fig3}b and~\ref{fig4}b with Fig.~\ref{fig5}). 
Escaping $\gamma$-ray spectra  
strongly depend on the viewing angle $\alpha$ not only in TeV range ($> 100$ GeV)
but also in the GeV range (see Fig.~\ref{fig5}). Moreover, the $\gamma$-ray fluxes 
expected in these two
energy ranges are strongly anticorrelated (large TeV flux is accompanied
by low GeV flux and the opposite).
This is due to the fact that at directions of the large optical depths the energy 
converted from primary electrons in the IC process to the TeV $\gamma$-rays is 
re-distributed to the GeV 
energies. In directions where the optical depths are low, the energy of TeV 
$\gamma$-rays is not so efficiently degraded in the cascade process to lower energies.
As a result, relatively large TeV fluxes are accompanied by small GeV fluxes. 

\subsection{LSI +61$^{\rm o}$ 303}

The binary system LSI +61$^{\rm o}$ 303 differs in some aspects from LS 5039.
Although both binaries contain massive stars with similar radii, the surface 
temperature of the star in LSI +61$^{\rm o}$ 303 is a factor of 
$\sim 0.7$ lower which results in the radiation energy density by a factor of $\sim 4$ 
lower. Also the distance of the compact object from the massive star in 
LSI +61$^{\rm o}$ 303 change in larger range, from $1.5r_\star$ up to $9.15r_\star$. 
Therefore, primary $\gamma$-rays injected in the jet propagate through the radiation field 
which vary with larger amplitude with full orbital period than in LS 5039. 
$\gamma$-ray spectra escaping toward the observer for the case of injection of 
primary $\gamma$-rays and electrons change more significantly with the observation angle 
$\alpha$. As an example, we show the $\gamma$-ray spectra for the case of injection
of primary $\gamma$-rays at the base of the jet, i.e. for $z = 0$, (Fig.~\ref{fig6}), 
and at the distance $z = 3r_\star$ from the base of the jet (Fig.~\ref{fig7}). 
$\gamma$-ray spectra for the periastron passage of the injection place are quite similar
to these ones expected from LS 5039 (compare the results in the upper Figs.~\ref{fig3}c 
and~\ref{fig4}c with the upper Figs.~\ref{fig6}c and~\ref{fig7}c). 
The effects of the lower surface temperature of the massive star
in LSI +61$^{\rm o}$ 303 and  the distance at the periastron passage almost compensate. 
However, at the apastron passage and at the base of the jet, the absorption dips 
(above $\sim 100$ GeV) are much lower in the case of LSI +61$^{\rm o}$ 303
(compare the bottom Figs.~\ref{fig6}c and~\ref{fig7}c with the bottom Figs.~\ref{fig3}c 
and~\ref{fig4}c). Even at the apastron passage of the compact object in 
LSI +61$^{\rm o}$ 303 and 
at the distance of injection place from the base of the jet $z = 3r_\star$,
$\gamma$-ray spectrum above $\sim 100$ GeV should still vary by a factor of 
$\sim 5$ with the angle $\alpha$ (see Fig.~\ref{fig7}c). 
Note that secondary cascade $\gamma$-ray spectra produced by primary $\gamma$-rays,
which are injected at large distance from the massive star (at the apastron passage
and at $z = 3r_\star$),
show strong cut-offs after $\sim 2$ TeV. This is the result of a relatively weak absorption
of primary $\gamma$-rays with energies above a few TeV, combined with the kinematics of
the IC scattering process for the case of strongly anisotropic radiation field of soft
photons from the massive star.

$\gamma$-ray spectra emerging from the binary system LSI +61$^{\rm o}$ 303
in the case of injection of primary electrons at the periastron distance are very 
similar to the spectra expected from LS 5039  
(Fig.~\ref{fig5}a,c with Fig.~\ref{fig8}a,c). The absorption features are only slightly 
stronger. $\gamma$-ray spectra, produced by electrons injected farther from the base of 
the jet, e.g. at the distance $z = 3r_\star$, are still strongly
influenced by the cascade effects showing strong dependence on the observation angle
in the GeV and TeV energy ranges (see Fig.~\ref{fig8}c). However, at the apastron 
distance, the $\gamma$-ray spectra look different. They show strong dependence
on the observation angle in the GeV energy range but relatively weak dependence
in TeV energy range (e.g. Fig.~\ref{fig8}b). $\gamma$-ray spectra produced at 
the apastron passage, but at distances from the base of the jet of the order of a few
stellar radii, do not depend strongly on $z$ (see Figs.~\ref{fig8}b and d),
since electrons injected at such distance still propagate in quite similar radiation 
field. In fact, for the apastron passage of the compact object in 
LSI +61$^{\rm o}$ 303 (at $9.15r_\star$), the distance  
of the injection place along the jet ($z = 3r_\star$) is located only 
at the distance of $9.63r_\star$ from the massive star (i.e very similar to the 
apastron distance). Therefore, TeV $\gamma$-ray fluxes observed from 
LSI +61$^{\rm o}$ 303 at the apastron passage
should be relatively larger in respect to their GeV fluxes in comparison to the
expectations from the massive binary LS 5039 in which the TeV fluxes should be
significantly lower than their corresponding  GeV fluxes. 

\section{Phase dependent gamma-ray spectra and light curves}

The observer is located differently in respect to the orbital plane of the binary 
systems LS 5039 and LSI +61$^{\rm o}$ 303. The inclination angles of 
the binary systems are probably quite similar, i.e 25$^{\rm o}$ in the case of 
LS 5039 (assuming that the compact object is a black hole)
and 30$^{\rm o}$ applied in the case of LSI +61$^{\rm o}$ 303. However the 
azimuthal angles of the observer, measured in respect to the periastron passage, 
are 225$^{\rm o}$ in LS 5039 and 70$^{\rm o}$ in LSI +61$^{\rm o}$ 303.
Therefore, phase dependent $\gamma$-ray spectra and $\gamma$-ray light curves expected 
from these two binary systems should have different features. 
Below we show the $\gamma$-ray light curves from these binaries at the  
GeV energies ($1-10$ GeV) and the TeV energies ($>100$ GeV), based on the 
calculations of the IC $e^\pm$ pair
cascades occurring inside the radiation field of their massive stars.
The phase dependent spectral features of the $\gamma$-ray emission are also 
discussed.

\begin{figure*}
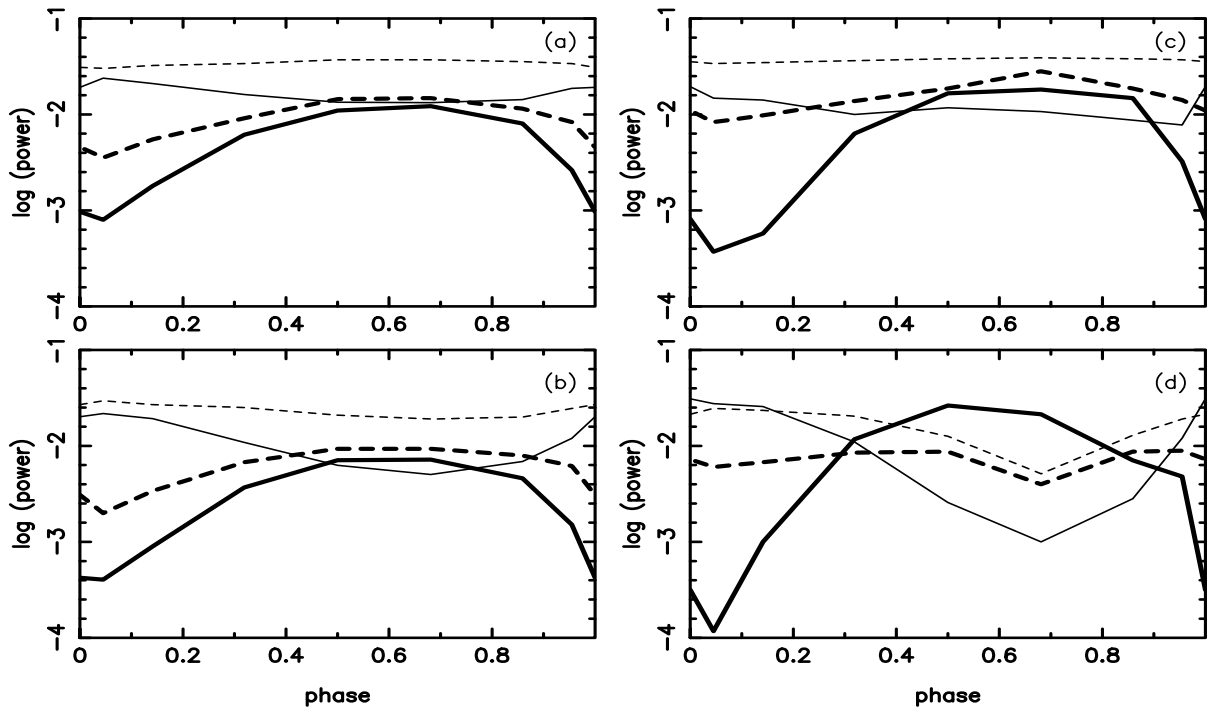

\vskip 9.5truecm
\includegraphics{gphase-obs.eps}
\includegraphics{ephase-obs.eps}
\includegraphics{gphase-obs-i60.eps}
\includegraphics{ephase-obs-i60.eps}
\caption{The power in $\gamma$-rays which escape from the binary system LS 5039
toward the observer (the $\gamma$-ray light curve) at the inclination angle 
$\theta= 25^{\rm o}$ ((a) and (b)) and $\theta= 60^{\rm o}$ ((c) and (d)), for 
the case of isotropic injection of primary $\gamma$-rays ((a) and (c)) and primary 
electrons ((b) and (d)),
as a function of the phase of the injection place of the primary particles
({\it phase} is the time measured from the periastron passage 
divided by the orbital period of the binary system) 
at photon energies in the range 1-10 GeV (thin curves) and above 100 GeV
(thick curves). The injection place of primary particles occurs in the
jet. The light curves are shown for the injection place at the base of the jet
$z\approx 0r_\star$ (full curves) and at the distance $5r_\star$ along the jet (dashed curve).}
\label{fig9}
\end{figure*}
\subsection{LS 5039}

We have calculated the $\gamma$-ray luminosities escaping toward 
the observer as a function of the phase of the injection place of primary particles
in the jet launched from the compact object in LS 5039. The light curves are obtained
for two energy ranges, the GeV range ($1-10$ GeV) and the TeV range ($>100$ GeV),
in the case of isotropic injection of 
primary $\gamma$-rays (Fig.~\ref{fig9}a and c) and primary electrons (Fig.~\ref{fig9}b 
and d), 
with the power law spectra and spectral index 2 (as discussed in Sect.~4.1). 
We consider the injection regions 
at the base of the jet, $z = 0$ (full curves) and at the distance $z = 5r_\star$ 
(dashed curves) along the jet and two inclination angles of the observer
$\theta= 25^{\rm o}$ (Fig. \ref{fig9}a and b) and $\theta= 60^{\rm o}$ (Fig. \ref{fig9}c and d). 
In order not to complicate the geometry too much, 
only the case of perpendicular propagation of the jet in respect to the plane of 
the binary system is considered. The calculations of the $\gamma$-ray spectra for 
the jets aligned at some angle to the plane of the binary system are straightforward.
They will be discussed in another work.

If the primary particles are injected at the base of 
the jet, then the $\gamma$-ray power in the GeV energy range vary by a factor of 
$\sim 2$ and in TeV energy range by a factor greater than $\sim 10$ for $\theta= 25^{\rm o}$ 
(see Fig.~\ref{fig9}a and b). The maximum in TeV $\gamma$-ray light curve  
occurs at the phase $\sim 0.6$ and corresponds to the minimum in the GeV $\gamma$-ray 
light curve. Therefore, based on these cascade calculations we predict clear 
anticorrelation between the GeV and TeV $\gamma$-ray fluxes. 
The level of variability of the GeV fluxes is larger in the case of injection of 
primary electrons in comparison to the injection of primary $\gamma$-rays since 
the primary (unabsorbed) $\gamma$-rays also 
contribute to the predicted total $\gamma$-ray light curve. Therefore,
GeV $\gamma$-ray light curve is quite smooth for the case of injection of primary $\gamma$-rays.
This feature might give some insight into the production mechanism
of primary particles ($\gamma$-rays or electrons) allowing to distinguish which 
particles are accelerated, hadrons (responsible for the primary $\gamma$-rays) or 
electrons?. 
If primary particles are injected farther from the base of the jet, then the
amplitude of the $\gamma$-ray emission drops in both energy ranges (see dashed curves
in Fig. \ref{fig9}a and b).
However, the level of variability is still larger in the case of injection of primary
electrons.  

The $\gamma$-ray light curves for the observer located at larger inclination angle,
i.e. $\theta= 60^{\rm o}$, show significantly larger variability than for $\theta= 25^{\rm o}$. 
For example, the $\gamma$-ray power can change by almost an order of magnitude in 
the GeV energy range and by two orders of magnitudes in the TeV energy range for the case
of injection of primary electrons (see Fig. \ref{fig9}d). 
Therefore, in principle the level of $\gamma$-ray variability might help to 
distinguish between different interpretations of the observational data concerning 
the inclination of the binary system in LS 5039 and put some insight on the nature of the compact object.

\begin{figure*}
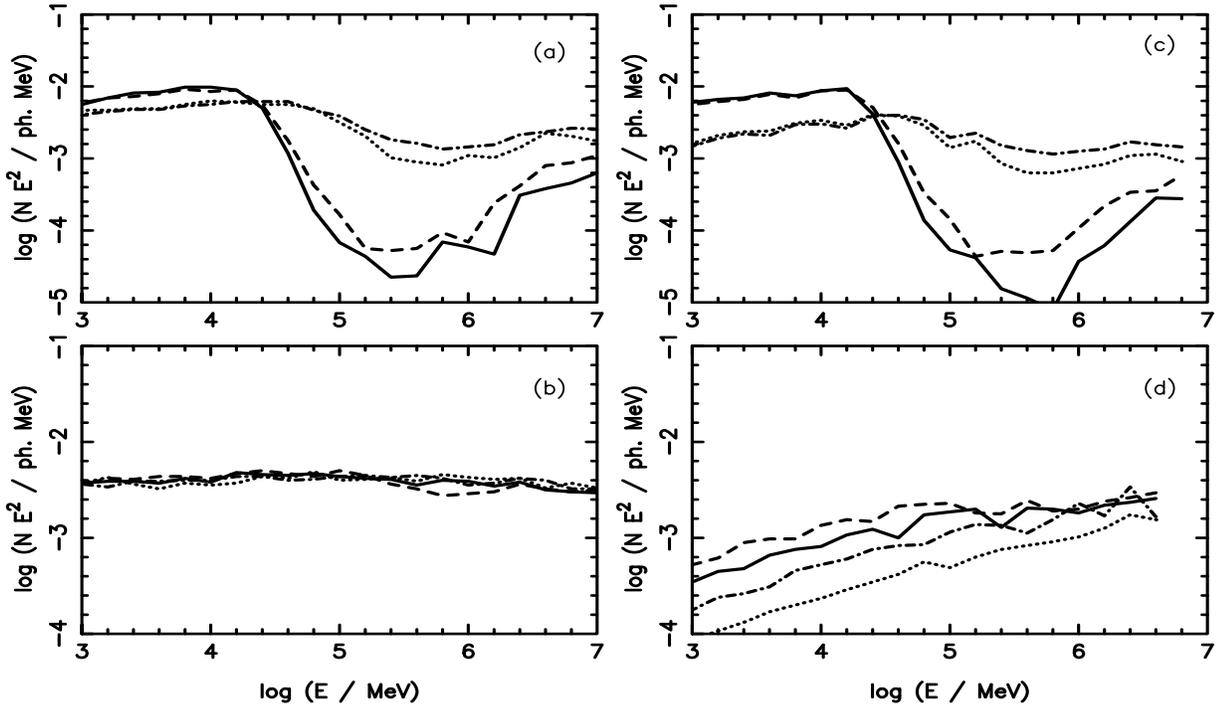

\vskip 9.5truecm
\includegraphics{specobs.eps}
\includegraphics{specobsz5.eps}
\includegraphics{especobs.eps}
\includegraphics{especobsz5.eps}
\caption{$\gamma$-ray spectra which escape from the binary system LS 5039
toward the observer at the inclination angle $\theta= 25^{\rm o}$ for different locations  
of the injection place of the source of primary particles 
((a) and (b) - $\gamma$-rays, and (c) and (d) - electrons)
defined by the azimuthal angle measured from the periastron: 
$\omega = 0^{\rm o}$ (full curve), 
$90^{\rm o}$ (dashed), $180^{\rm o}$ (dot-dashed), and $270^{\rm o}$ (dotted),
and at the distance from the base of the jet 
$z\approx 0r_\star$ ((a) and (c)) and $z = 5r_\star$ (figures (b) and (d)).}
\label{fig10}
\end{figure*}
\begin{figure*}
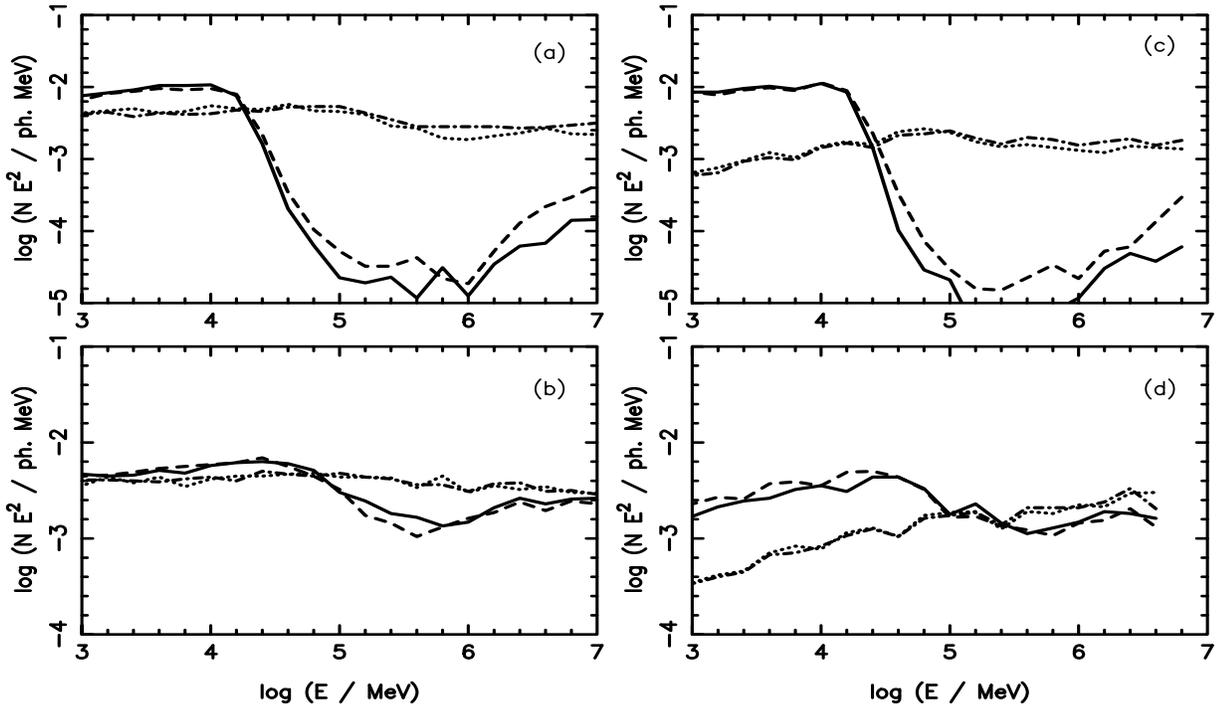

\vskip 9.5truecm
\includegraphics{specobs-i60.eps}
\includegraphics{specobsz5-i60.eps}
\includegraphics{especobs-i60.eps}
\includegraphics{especobsz5-i60.eps}
\caption{As in Fig.~\ref{fig10} but for the inclination angle of the binary system
$\theta= 60^{\rm o}$.}
\label{fig10a}
\end{figure*}

In Fig.~\ref{fig10} and ~\ref{fig10a}, we show the $\gamma$-ray spectra which escape to 
the observer for a few selected phases (time from the periastron divided by the orbital 
period) of the compact object in all considered 
above scenarios for the case of the inclination of the binary system equal to 
$\theta= 25^{\rm o}$ and $60^{\rm o}$, respectively.
The cascade $\gamma$-ray spectra produced by primary particles injected at the base of
the jet drop
suddenly above a few tens of GeV, reach the minimum at a few 100 GeV and becomes
flat (spectral index $<$2) at higher energies. Between $\sim 0.1-1$ TeV, the spectral 
index does not change significantly with the phase of the injection place. 
The TeV $\gamma$-ray deficit is larger for the injection of primary electrons and for
larger inclination angles. However, the spectral indexes in the GeV and TeV energies
are close to two in spite of such large difference in the level of emission.
These features are generally consistent with the observations of LS 5039
in the GeV and TeV energy ranges.
The spectra look completely different for the injection place farther along the 
jet. For example, already at the distance of $5r_\star$ from the base of the jet
and $\theta= 25^{\rm o}$, the
total cascade $\gamma$-ray spectra are almost independent on the phase of the compact 
object in the case of injection of primary $\gamma$-rays (see Fig.~\ref{fig10}b). 
However, for larger inclination angles of the binary, e.g $\theta= 60^{\rm o}$, small absorption feature starts
to appear at TeV energies when the compact object is at the periastron.
In the case of injection of primary electrons, the TeV $\gamma$-ray spectra
are on similar level but differ significantly in the GeV energies. This GeV spectrum
is flatter than in the case of injection of primary $\gamma$-rays 
(spectral index close to $\sim 1.5$ versus a slightly flatter than 2), due to 
the lack of domination of the GeV spectrum by the primary $\gamma$-rays. 
These weak dependences of the $\gamma$-ray spectra in the TeV energies are due to small 
differences in the
angles between the observer and the injection place of the primary particles farther 
from the base of the jet (e.g. considered here $z = 5r_\star$)
for different phases of the binary system. 
Note however, that these angles show stronger differences for larger inclination angles 
of the binary system (compare Fig.~\ref{fig10} and Fig.~\ref{fig10a}).

\subsection{LSI +61$^{\rm o}$ 303}

\begin{figure}
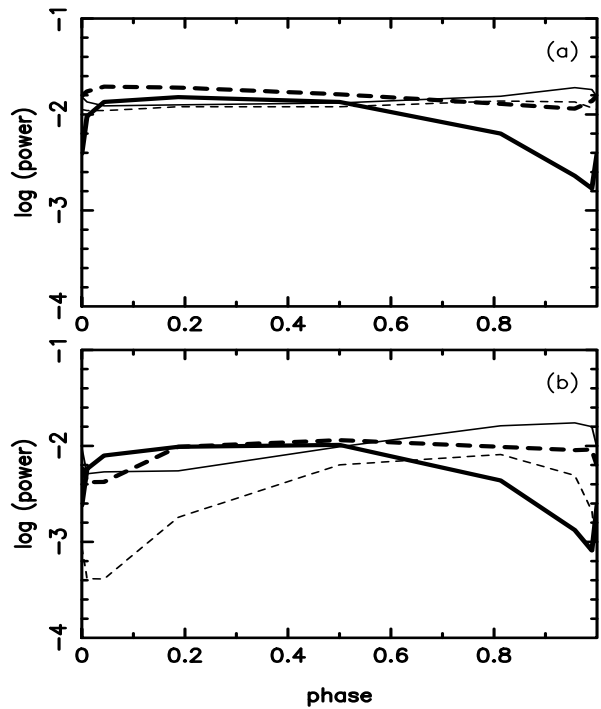

\vskip 9.5truecm
\includegraphics{gphaselsi.eps}
\includegraphics{ephaselsi.eps}
\caption{As in Fig.~\ref{fig9} but for the binary system LSI +61$^{\rm o}$ 303
(fi=or the inclination $\theta= 30^{\rm o}$).
Primary $\gamma$-rays (figure a) and primary electrons (b) are 
injected at the distance from the base of the jet 
$z\approx 0r_\star$ (full curves) and $z\approx 3r_\star$ (dashed curves).}
\label{fig11}
\end{figure}

\begin{figure*}
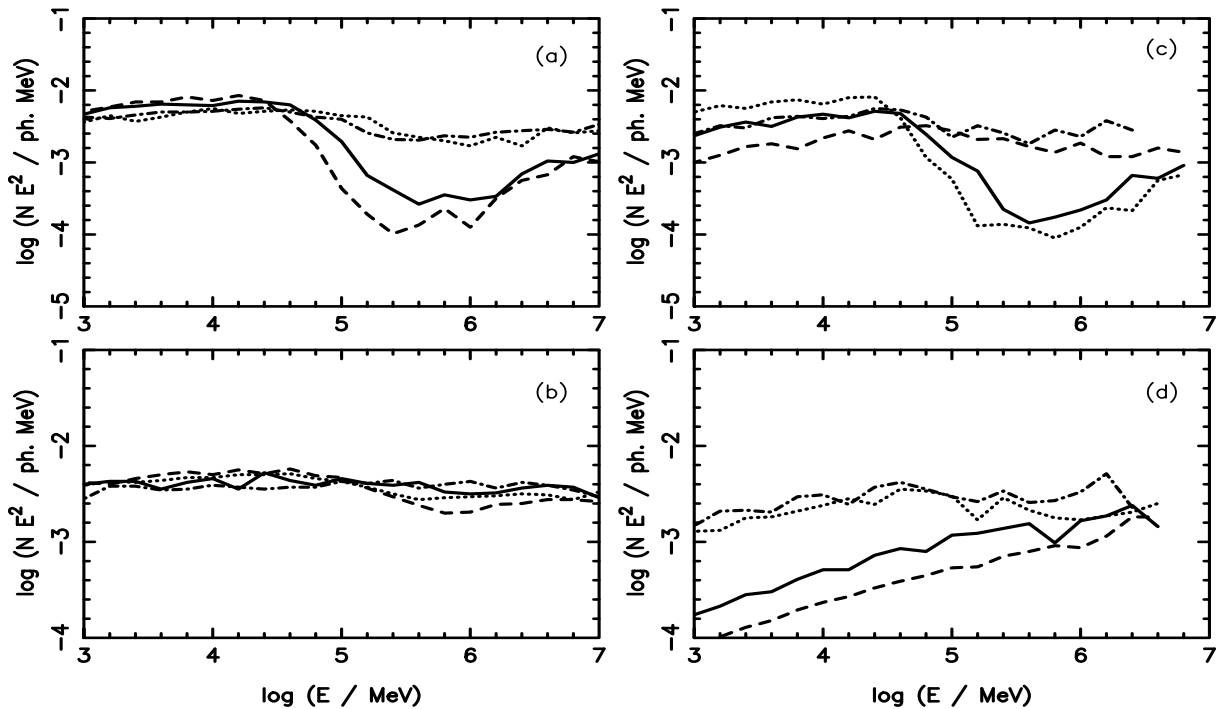

\vskip 9.5truecm
\includegraphics{specobslsi.eps}
\includegraphics{specobslsiz3.eps}
\includegraphics{especobslsi.eps}
\includegraphics{especobslsiz3.eps}
\caption{As in Fig.~\ref{fig10} but for LSI +61$^{\rm o}$ 303.
Primary $\gamma$-rays (figures (a) and (b)) and primary electrons (figures
(c) and (d)) are injected at the distance from the base of the jet 
$z\approx 0r_\star$ (figures (a) and (c)) and $z = 3r_\star$ 
(figures (b) and (d)).}
\label{fig12}
\end{figure*}

The features of the $\gamma$-ray light curves and phase dependent spectra 
expected from the binary system LSI +61$^{\rm o}$ 303
are qualitatively similar to these ones discussed above for the binary system 
LS 5039 (see Figs.~\ref{fig11} 
and ~\ref{fig12}). However, there also some significant differences. 
For example, the maximum in TeV $\gamma$-ray light curve of LSI +61$^{\rm o}$ 303
is broader and appears at 
the phase $\sim$0.2-0.5. The level of variability at TeV energies is smaller
(variability by a factor less than $\sim 10$). 
The anticorrelation between the GeV and the TeV fluxes is not so excellent as in LS 5039
(possible shift in phase by $\sim$0.2). Significantly smaller variability is expected in the GeV 
and TeV energy ranges, if the 
primary $\gamma$-rays are injected already at the distance of $3r_\star$ from the base 
of the jet. The level of variability in the GeV range is larger 
in respect to the TeV energy range when compared to the results obtained fro LS 5039 in 
the case of the primary electrons injected at the distance of $3r_\star$ from the base of the jet. 
The absorption dips in the TeV $\gamma$-ray energy range are shallower than expected in LS 5039.

In summary, propagation effects influence in similar way the
$\gamma$-ray spectra escaping from the binary systems LSI +61$^{\rm o}$ 303 
and LS 5039 when the injection place of the primary particles is relatively close
to the massive stars (not far from the periastron passage and for small distances from 
the base of the jet). However, significant differences are expected for the injection of
primary electrons farther along the jet and at the apastron passage as shown in 
Figs. \ref{fig11} and \ref{fig12}.

\section{Conclusions}

We have performed Monte Carlo simulations of the propagation of high energy
$\gamma$-rays inside compact massive binaries of the microquasar type, applying as 
an example parameters of LS 5039 (recently observed in  TeV $\gamma$-rays and
suggested as a counterpart of the EGRET source) and  LSI +61$^{\rm o}$ 303 
(similar parameters to LS 5039 and also suggested as a counterpart of the EGRET 
source). In both sources the optical depths for TeV $\gamma$-rays are greater than unity
for most of the propagation directions, provided that the injection place of primary
particles ($\gamma$-rays or electrons) is relatively close to the base of the jet 
launched from the inner part of the accretion disks around compact objects. 
Therefore,
these primary particles initiate IC $e^\pm$ pair cascades in the anisotropic radiation
of the massive stars. We have calculated the $\gamma$-ray spectra produced in such 
cascades and investigated their basic features. 
It is concluded that the absorption dips in TeV $\gamma$-ray spectra emerging 
toward the observer are not so drastic as predicted in the earlier papers, e.g. by 
B\"ottcher \& Dermer~(2005) and  Dubus~(2005). This is clearly seen by    
comparing $\gamma$-ray spectra calculated with pure absorption (e.g 
Figs.~\ref{fig3}a, 
\ref{fig4}a, \ref{fig6}a, and \ref{fig7}a) with the corresponding total $\gamma$-ray
spectra produced in the cascade processes inside the binary system  
(Figs.~\ref{fig3}c, \ref{fig4}c, \ref{fig6}c, and \ref{fig7}c).    
This less pronounced dips are due to the re-distribution of energy in the primary spectrum
of $\gamma$-rays from the region above $\sim 1$ TeV  to $0.1-1$ TeV energy range.
Since the soft radiation field created by the massive stars in these two binaries
are quite similar during periastron passages of their compact objects, 
the $\gamma$-ray spectra produced in the cascade processes do not differ significantly
in the case of LS 5039 and LSI +61$^{\rm o}$ 303. However, the propagation effects are
responsible for essential differences at the apastron passage, since 
LSI +61$^{\rm o}$ 303
is more extended (the apastron distance in LSI +61$^{\rm o}$ 303 is
a factor of $\sim$2 larger than in LS 5039).

Let us concentrate at first on the features of $\gamma$-ray emission from LS 5039.
The $\gamma$-ray luminosity from this source observed in the GeV energy range 
(assuming that the identification with the EGRET source by Paredes et al.~2000
is correct) 
is about two orders of magnitude larger than observed in TeV energy range
(Aharonian et al.~2005b). However, spectra reported in these two energy ranges 
are quite similar (spectral index close to 2). 
These spectral features can be naturally explained by the propagation effects considered 
in this paper, i.e. IC e$^\pm$ pair cascading processes, inside the binary system. 
The primary $\gamma$-rays or electrons could be injected with a simple power law 
spectrum (and spectral index close to 2), relatively close to the base of the jet. 
The $\gamma$-ray luminosity expected in this propagation model at GeV energies 
($1-10$ GeV) should vary by a factor of $\sim 2-3$, 
and at TeV energies ($>$100 GeV) by a factor of $\sim 10$ with the orbital period of 
the binary system LS 5039. In fact, possible variation of the TeV signal
by a factor of $\sim 3$ has been recently suggested by 
Casares et al. (2005), although not statistically significant 
(Aharonian et al. 2005b).
Moreover, we show that, due to the propagation (IC $e^\pm$ pair cascade) effects,
the maximum in TeV $\gamma$-ray light curve should occur at phase $\sim 0.6$
(measured as an azimuthal angle from the periastron passage, see Fig.~\ref{fig1}).
This seems to be inconsistent with the recent suggestion by Casares et al.~(2005b)
who, reanalyzed the HESS data with the new orbital parameters and, concluded that
maximum of TeV emission occurs at the phase $\sim 0.9$. 
This discrepancy does not allow us to put any definitive conclusions since the error 
bars in TeV $\gamma$-ray light curve, shown by Casares et al.~(2005b), 
are very large. However, if real, it can give some information on the efficiency
of acceleration process of the primary particles occurring in the jet 
(possibly linked with the efficiency of the accretion process) with the phase of 
the compact object on its orbit around the massive star. 
 
Based on the propagation calculations it is concluded that $\gamma$-ray spectra 
produced in such IC $e^\pm$ pair cascades should show 
clear anticorrelation between the fluxes in the GeV and TeV energy ranges (see the
light curves in Fig.~\ref{fig9}).
We also investigated the dependence of the propagation effects on the 
distance of the injection place from the base of the jet.
If primary particles are injected 
already at the distance of a few stellar radii from the base of the jet
(the case of $z = 5r_\star$ is discussed), 
then $\gamma$-ray fluxes expected from LS 5039 do not vary strong enough 
to explain relative luminosities in the GeV and TeV energy ranges.
Either a more complex shape for the spectrum of primary particles injected into the jet 
is required, e.g composed of at least three different power laws, or two different 
populations of primary particles are needed, or the origin of different parts of
$\gamma$-ray spectra in different regions of the jet has to be postulated.
Moreover, the shape of the $\gamma$-ray spectra escaping toward the observer depends also
on the inclination angle of the binary system which is not at present well known
(estimated on $25^{\rm o}$ for the case of a solar mass black hole and on $60^{\rm o}$ 
for the case of a neutron star). The level of variability of the TeV emission for the injection place
of primary particles at the base of the jet is clearly larger for the inclination angle $\theta= 60^{\rm o}$
than for $\theta= 25^{\rm o}$. However these differences seems to be to low in order
to put constraints on the inclination angle of binary system LS 5039 based only on the observations with
the present sensitivity Cherenkov telescopes.

If the identification of the binary system LSI +61$^{\rm o}$ 303 with the EGRET source
3EG J0241+6103 is correct (see introduction), then average $\gamma$-ray luminosity 
($>100$ MeV), 
$\sim 8\times 10^{34}$ erg s$^{-1}$, is only about a factor of three lower than
in the case of LS 5039. This suggests that $\gamma$-ray properties of 
these two sources should be quite similar. Our simulations of the propagation 
of primary $\gamma$-rays show that absorption of $\gamma$-rays 
is similar in 
both binaries at the periastron passage of the compact object (if primary particles 
are injected at the base of the jet)
but is less important at the apastron passage of LSI +61$^{\rm o}$ 303. 
The TeV fluxes from  LSI +61$^{\rm o}$ 303 predicted by the propagation effects should 
be relatively larger in respect to its GeV fluxes than 
observed in the binary system LS 5039. This feature might help in detection of 
LSI +61$^{\rm o}$ 303 at TeV energies. The maximum in TeV $\gamma$-ray light curve
predicted by the propagation effects in LSI +61$^{\rm o}$ 303 should occur at the phase $\sim 0.2-0.4$, 
measured from the periastron. The TeV $\gamma$-ray signal is expected to be modulated 
by an order of magnitude with the maximum
for the case of injection of primary electrons injected farther along the jet 
(at  a distance of a few stellar radii, see the case of $z = 3r_\star$ in 
Fig. \ref{fig11}). Also strong variability of the GeV flux is possible
(even by an order of magnitude) for the case of injection of primary electrons farther
from the base of the jet (see e.g. the case of $z = 3r_\star$ in Fig. \ref{fig11}b).
Injection of electrons at the base of the jet predict modulation by a factor of $\sim 2$
with the minimum when the compact object is in front of the massive star.
Re-analysis of the EGRET data indicates probable modulation of the GeV signal from 
the source toward LSI +61$^{\rm o}$ 303 (Tavani et al.~1998, Wallace et al.~2000) with the maximum
emission near periastron (Massi 2004). This is inconsistent with the predictions of 
studied here propagation effects for the case of injection of primary electrons
which show deficit of GeV emission close to periastron.
Therefore, if this modulation is real then the mechanism responsible for such modulation
of GeV emission should be even much more efficient since considered here propagation effects 
work in opposite direction. For the case of isotropic injection of primary $\gamma$-rays, the GeV signal is
only weakly modulated with the period of the binary system due to the domination of primary $\gamma$-rays
at such low energies.

In this paper we have only analyzed the effects of propagation of $\gamma$-rays inside these two 
binary systems assuming that efficiency of particle acceleration and spectra of 
injected particles do not depend on the location of the injection place of primary 
particles along the jet, i.e. they are uniformly distributed along the  
jet, i.e. from $z = 0r_\star$ to at $z = 5r_\star$ (LS 5039) or $3r_\star$ 
(LSI +61$^{\rm o}$ 303). Proper analysis of $\gamma$-ray emission from these
sources require, accept considering the IC $e^\pm$ pair cascade effects, 
a detailed model for efficiency of accretion process,
conversion of accretion energy into particles, acceleration of particles 
(their spectra), and production of $\gamma$-rays.
Such models, which unfortunately do not take into account the cascade effects 
analyzed in this paper but discuss only production of $\gamma$-rays in the 
Inverse Compton process far away from the base of the jet, have been recently 
discussed by Bosch-Ramon, Romero \& Paredes~(2005) and Dermer \& B\"ottcher~(2005).

\section*{Acknowledgments}
I would like to thank the referee, Dr G. Dubus, for many useful comments. 
This work is supported by the Polish MNiI grant No. 1P03D01028 and the KBN grant 
PBZ KBN 054/P03/2001. 


\end{document}